\documentclass[12pt,english]{article}

\usepackage[T1]{fontenc}
\usepackage[latin9]{inputenc}
\usepackage{color}
\usepackage{babel}

\usepackage{amsmath}
\usepackage{amsthm}
\usepackage{caption}
\usepackage{graphicx}
\usepackage[authoryear]{natbib}
\PassOptionsToPackage{normalem}{ulem}
\usepackage{ulem}
\usepackage[unicode=true,pdfusetitle,
 bookmarks=true,bookmarksnumbered=false,bookmarksopen=false,
 breaklinks=false,pdfborder={0 0 0},pdfborderstyle={},backref=false,colorlinks=false]
 {hyperref}

\makeatletter
\theoremstyle{plain}
\newtheorem{thm}{\protect\theoremname}
\theoremstyle{plain}
\newtheorem{prop}[thm]{\protect\propositionname}
\newtheorem{definition}{Definition}

\usepackage{amsmath}
\usepackage{fullpage}
\usepackage{booktabs}
\usepackage{xcolor}

\DeclareMathOperator{\Tr}{Tr}
\DeclareMathOperator{\Trace}{Trace}
\DeclareMathOperator*{\argmin}{arg\,min}
\DeclareMathOperator*{\argmax}{arg\,max}
\usepackage{bbm}

\newcommand\mysmalleq[2]{{#1}\mkern-0.2mu{=}\mkern-0.2mu{#2}}

\makeatother

\providecommand{\propositionname}{Proposition}
\providecommand{\theoremname}{Theorem}

\begin{document}
\title{Optimal index insurance and basis risk decomposition:\\
an application to Kenya}
%% REVIEW 
\author{Matthieu Stigler, David Lobell\\
Center on Food Security and the Environment, Stanford University}
%\author{}
\date{This version: November 1 2021. For a more recent version, see the published article in \href{https://doi.org/10.1111/ajae.12375}{American Journal of Agricultural Economics}}
%% REVIEW \date{Version: \today\\
%Latest version available \textcolor{blue}{\uline{\href{https://drive.google.com/file/d/1dr7oR0vlfi8HXh5VEEoqz9EcC4aEkO6F/view?usp=sharing}{here}}}
%PRELIMINARY, do NOT CIRCULATE
%}

%\begin{titlingpage} %% AJAE
\maketitle
%\linenumbers % AJAE

% \vspace{-3.5cm} %AJAE

\begin{abstract}

Index insurance is a promising tool to reduce the risk faced by farmers, but high basis risk, which arises from imperfect correlation between the index
and individual farm yields, has limited its adoption to date. Basis risk arises from two fundamental sources: the intrinsic heterogeneity within an insurance zone (\emph{zonal risk}), and the lack of predictive accuracy of the index (\emph{design risk}). Whereas previous work has focused almost exclusively on design risk, a theoretical and empirical understanding of the role of zonal risk is still lacking. 

Here we investigate the relative roles of zonal and design risk, using the case of maize yields in Kenya. Our first contribution is to derive a formal decomposition of basis risk, providing a simple upper bound on the insurable basis risk that \emph{any} index can reach within a given zone. 
Our second contribution is to provide the first large-scale empirical analysis of the extent of zonal versus design risk. To do so, we use satellite estimates of yields at 10m resolution across Kenya, and investigate the effect of using smaller zones versus using different indices. Our results show a strong local heterogeneity in yields, underscoring the challenge of implementing index insurance in smallholder systems, and the potential benefits of low-cost yield measurement approaches that can enable more local definitions of insurance zones. 

Keywords: index insurance, agricultural risk management basis risk, risk decomposition, satellite data

JEL codes: G22, O16, Q14
\end{abstract}

% \end{titlingpage} % AJAE

% \linenumbers % AJAE

\section{Introduction}

Risk is ubiquitous in agriculture and can have detrimental effects on producers, particularly in smallholder agriculture where access to formal credit and savings is limited. Index insurance, where payouts are triggered based on an external index, has been advocated as a promising tool to reduce risk compared to traditional individual-based insurance. By de-linking payments from individual loss, index insurance removes moral hazard and reduces adverse selection, the two main issues that plague individual-based insurance. By relying on measuring an index at an aggregate level instead of measuring each individual field, it furthermore reduces transaction costs. 

Despite its theoretical benefits, success of index insurance has been rather limited, with a low take-up for most products \citep{ColeGineEtAl2013,ColeGineEtAl2017}. A core reason for this low success lies in the concept itself of index insurance: by de-linking payment from individual losses, index insurance introduces \emph{basis risk}, the possibility that a farmer suffers a harvest loss and yet does not receive any payment. As pointed out by \citet{Clarke2016}, this eventuality suggests that a highly risk-averse farmer would not want to purchase index insurance. Unfortunately, the basis risk appears to be high in practice. \citet{ClarkeMahulEtAl2012} report correlations of only 0.14 between payments and yields in India, predicting a probability as high as 1/3 of not receiving any indemnity even in the catastrophic case of yields falling to zero. \citet{JensenBarrettEtAl2016} likewise find substantial basis risk in a livestock insurance in Kenya. 

The presence of basis risk has triggered a large literature, extending now well beyond the economic realm, on a quest for the ``perfect'' index. Indices considered range from input-based indices such as precipitation, temperatures, soil moisture \citep{VroegeBucheliEtAl2021} or satellite-derived vegetation indices
\citep{ChantaratMudeEtAl2013} to output-based indices using average yields, either from official statistics \citep{SkeesBlackEtAl1997}, sampling surveys \citep{WB_Clarke_Kenya_2015} or from satellite estimates \citep{FlatnesCarterEtAl2018}; see \citet{BenamiJinEtAl2021} for a survey. 
With the proliferation of these studies,
there is some confusion with the notion of basis risk. In some studies, finding the perfect index is sometimes conflated with the task of finding the best predictor of individual or aggregate yields. Implicitly assumed is the idea that if one were able to perfectly predict yields, one would obtain a perfect index without any basis risk. This neglects, however, the important fact that index insurance is still using a regional-based index. Unless yields are perfectly correlated, any regional index will be an imperfect predictor of individual farm-level yields and hence will introduce basis risk. However, there is no clear understanding in the literature on the extent of this \emph{irreducible} basis risk that simply arises from using a regional index. As a result, it is currently impossible to distinguish index quality from zone quality: a finding of high total basis risk can correspond to both a ``good index within a bad zone'' or to a ``bad index within a good zone''. 

\begin{figure}

\begin{centering}
%\centering{}
\includegraphics[width=0.95\columnwidth]{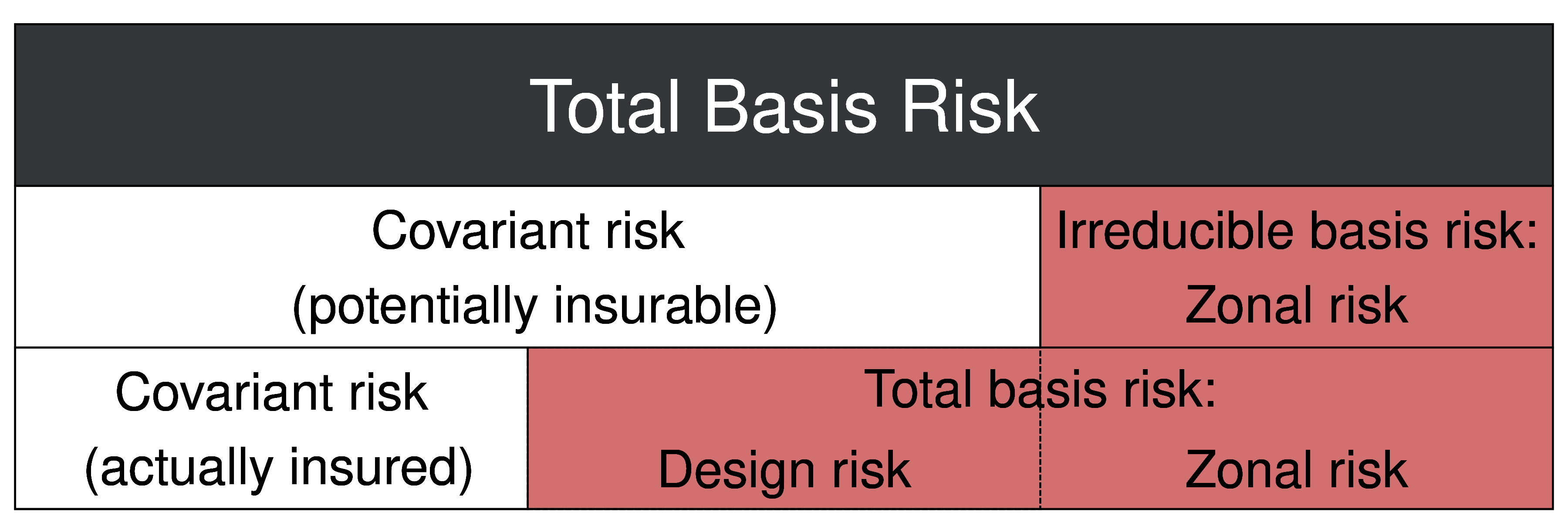}
\end{centering}

\caption{Decomposition of basis risk\label{fig:Decomposition-of-basis-risk-Elabed}.  \textmd{Source: adapted from \citet{BenamiJinEtAl2021}.}}
\end{figure}

In this paper, we seek to refocus the debate on basis risk by examining the role of irreducible zonal risk. To do so, we formalize the decomposition of basis risk advocated by \citet{ElabedBellemareEtAl2013}. According to these authors, basis risk arises from two distinct sources, 1) the \emph{zonal} risk\footnote{\citet{ElabedBellemareEtAl2013} originally used the term
of \emph{idiosyncratic} risk to denote what we call here \emph{zonal} risk. We prefer the term \emph{zonal} as we feel that it conveys better the fact that this risk only stems from using the same index for a zone. Using a zone with only one field would effectively bring this risk to zero. } due to the use of a regional index to predict individual yields, and 2) the \emph{design} risk due to lack of accuracy of the regional index at summarizing the aggregate covariant movement in yields. Intuitively, zonal risk is intimately
linked to the heterogeneity within a zone. The lower the correlation between fields, the higher the zonal risk. Conversely, the design risk stems from the use of a sub-optimal index that does not summarize well the covariant risk. Figure~\ref{fig:Decomposition-of-basis-risk-Elabed}, inspired by \citet{BenamiJinEtAl2021}, illustrates the decomposition of risk.
While a valuable and intuitive decomposition, \citet{ElabedBellemareEtAl2013} did not specify what the optimal index is. Specifying the optimal index is, however, a necessary condition for the decomposition to be valid: absent this specification, the distinction between zonal and design risk becomes arbitrary. Some authors assume that the optimal index is the zone-average yield \citep{FlatnesCarterEtAl2018} or an abstract weather-based index \citep{ConradtFingerEtAl2015,YuVandeveerEtAl2019}.
As we will show, this is not correct and can lead to nonsensical situations where an ``imperfect'' index performs better than a supposedly ``perfect'' one. 

Our first contribution is to derive the optimal index according to two correlation-based metrics of basis risk, both linked to the $R_{i}^{2}$ from the regression between each individual field $i$ and the index. Our first zone-specific metric $\overline{R^{2}}$ is the average of the $R_{i}^{2}$, while the second $\overline{\overline{R^{2}}}$ corresponds to a weighted-average of the same $R_{i}^{2}$. We derive the optimal index for each  metric, which provides a formal definition of zonal risk, giving an upper bound on the $\overline{R^{2}}$ \emph{any} index could reach for a given zone. Beyond this value, improvements in the $\overline{R^{2}}$ can only be achieved by re-designing the zone itself. Equipped with this upper bound, a practitioner can now distinguish index quality from zone quality and set priorities accordingly: if the upper bound is high, the zone is relatively homogeneous and it is worth embarking on the quest for a better index. On the other hand, if that maximum $\overline{R^{2}}$ turns out to be low, the quest for the best index is rather futile, and efforts should be rather spent on finding more
homogeneous (and likely smaller) zones. 

To compare our optimality theory from previous results in the literature, it is useful to emphasize the conditions under which our optimality result is obtained. We define here as index insurance a scheme that uses the same index and indemnity function over all fields, excluding insurance schemes with individual-specific indices or indemnities. Most of the optimality results in the literature take the opposite route, seeking to derive the optimal indemnity function for a given index. \citet{Mahul1999} derives the optimal field-specific indemnity function assuming that the index is the zone-average yield, and \citet{ZhangTanEtAl2019} extend it to a weather-based index. \citet{BourgeonChambers2003} extend \citet{Mahul1999}'s results
to include asymmetric information. \citet{MahulWright2003} consider the case of revenue insurance, and \citet{Mahul2001Climatic} of a weather-based index. Unlike these studies, we focus on the optimal
index regardless of the indemnity function, instead of the optimal indemnity for a given index. It is likely nonetheless that most of
the results from this strand of literature hold with the optimal index we derive.

A second way in which our approach differs from the literature is in the choice of the loss function and the focus on basis risk measures. Broadly speaking, two types of measures are used for index insurance: 1) basis risk metrics, which capture the relationship between individual yields and the index, and 2)  insurance benefits metrics, which evaluate the expected utility of yields with and without insurance. 
We focus here on a simpler basis risk measure based on the linear correlation between yields and the index. This is made mainly for convenience, as considering instead the correlation between ``post-indemnity'' yields and the index, or considering an expected utility loss function, makes the problem highly non-linear and hence hard to solve analytically. On the other hand, our metric has the advantage of being free of assumptions on the specific forms of the utility and indemnity functions. Furthermore, as shown in \citet{StiglerLobell_cropUS_wp}, the simple metric based on the average $R^{2}$ between yields and the index is a good predictor of the expected utility of index insurance for maize and soybeans in the United States. We observe a similar result in our robustness check section, finding a high correlation between our linear measure of basis risk and more sophisticated metrics.

Putting theory into practice, our second contribution is to showcase how satellite data can be used to measure the extent of zonal and design risk. We focus on Kenya, where the government has been investigating options for providing large-scale insurance to farmers, and where the \citet{WB_Clarke_Kenya_2015} recommends an index insurance scheme based on zone-average yields (also called area-yield insurance). Yet, the question of which and how many zones to use for the insurance remains open. Furthermore, at
the time of the report, the only available data were district-level averages, which are not informative about the  critical statistic for index insurance, within-district heterogeneity. To overcome these limitations, we use a dataset of yield estimates at a very fine (10m) resolution based on satellite data spanning the entire area cropped with maize in Kenya over 2016-2019 \citep{LobellEtAl2015,BurkeLobell2017,JinAzzariEtAl2017,JinAzzariEtAl2019}. Using  such a high-resolution dataset of yields allows us to go beyond the previous literature that used satellite data  to build indices but relied on ground-collected data to measure basis risk and insurance efficacy (see \citealp{ChantaratMudeEtAl2013,JensenBarrettEtAl2016,JensenStoefflerEtAl2019}).
Unlike these studies, we use the satellite data to measure basis risk, providing a detailed picture of within-district heterogeneity over many districts spread over multiple agro-ecological zones. Facing no restriction due to a ground-collected sample, we can further investigate the effect of varying the size of the insurance zone using various administrative boundaries, providing guidance for the implementation of the government's large-scale program. 

The paper is organized as follows: Section~\ref{sec:Theory} develops the theory of the optimal index, Section~\ref{sec:Data} describes the data used, Section~\ref{sec:Results} the main results, Section~\ref{sec:robustness} robustness checks, while Section~\ref{sec:Conclusion} concludes.

\section{Theoretical results\label{sec:Theory}}

In this section, we give a formal definition of the basis risk decomposition, define a measure of basis risk and derive the optimal index relative to the selected basis risk metric. Following the intuition in \citet{ElabedBellemareEtAl2013}, we construct the basis risk decomposition as:

\begin{definition}\label{def:decomposition} Let $BR(f)$ be a measure of basis risk associated with index $f$, with higher values  $BR(f')>BR(f)$ indicating that index $f'$ has more basis risk than index $f$.  Define as $f^\star$ the optimal index that minimizes $BR(\cdot)$, i.e. $f^\star \equiv \argmin_f BR(f)$. Then we define:
\begin{itemize}
    \item Total basis risk: $BR(f)$
    \item Zonal risk: $BR(f^\star)$. 
    \item Design risk: $BR(f) - BR(f^\star)$
\end{itemize}
\end{definition}

The \emph{zonal} risk $BR(f^\star)$ corresponds to the minimum irreducible basis risk attainable within a zone. It arises due to the fact that we are using a regional index to predict individual yields. The \emph{design} risk  is then defined as the difference between the optimal and actual index $BR(f) - BR(f^\star)$ and arises due to the use of a sub-optimal index. Together, these two components form the \emph{total} basis risk $BR(f)$, which is the value measurable by a practitioner when evaluating the basis risk of a given index. 

To be valid, the basis risk decomposition requires that $BR(f) \geq BR(f^\star) \: \forall f$. If this is not the case, design risk can become negative, leading to a nonsensical situation where design risk reduces total basis risk. This implies that the decomposition cannot hold for individual-level basis risk $BR_i(\cdot)$: for a given index $f$, one can always generate a negative design risk on a single field by trivially setting the index to that field's own yield, $f' =y_i$. In other words, there is no index $f^\star$ that  satisfies  $BR_i(f) \geq BR_i(f^\star)$  simultaneously $\forall f$ and $\forall i$. This makes it clear that the basis risk decomposition and the concept of design risk can only hold at the zone level, for a zone-specific basis risk measure $BR=f(BR_i)$.

We now introduce some notation to define our zone-wide basis risk metric $BR$. Let us denote by $y_{it}$ the yields for field $i=1\ldots N$ at time $t=1\ldots T$, and by $f_{t}$ the index. The index can be any input- or output-based measure, with the index by definition taking a constant value across space within the same zone, i.e $f_{it}=f_{t}\:\forall i$. We distinguish two types of indices, the output-based ones using a weighted average of observed yields, $f_{t}^{Y}\equiv\sum_{i=1}^{N}w_{i}y_{it}$ and the input-based ones using a variable x, $f_{t}^{X}\equiv\sum_{i=1}^{N}w_{i}x_{it}$. Typical output-based indices are the zone-average yields, $f_{t}^{Y}=\bar{y}_{t}$, and typical input-based indices rely on precipitation or other weather variables. In matrix notation, denoting by $Y$ the $T\times N$ matrix of yields, an output-based index can be written as $f^{Y}=Yw$, where $w$ is a $N\times1$
vector of weights. 

\citet{Miranda1991}'s seminal paper on the benefits of index insurance assumed an index based on zone-average yields:
%\footnote{Centering by $\bar{\bar{y}}_{\cdot\cdot}\equiv 1/T \sum \bar{y}_t$ as in Miranda does not change $\beta$.}

\begin{equation}
y_{it}=\alpha_{i}+\beta_{i}\bar{y}_{t}+\epsilon_{it}\label{eq:Mir_reg}
\end{equation}

Interpreting this model in a general way, we assume that yields follow an arbitrary distribution with covariance matrix $\Sigma$ of dimension $N\times N$. A diagonal element of $\Sigma$ represents field's $i$ variance, $\sigma_{i}^2$, while a typical off-diagonal element $\sigma_{ij}$ represents the covariance between field $i$ and $j$. Under this interpretation, $\beta_i$ is the least-square projection of $y_{it}$ on $\bar{y}_{t}$ from \eqref{eq:Mir_reg}. Denoting by $\mathbf{1}$ a vector of ones of dimension $N\times 1$, the vector of $\beta$ from regression \eqref{eq:Mir_reg} using the zone average $f=\bar{y}$ as index can be written in compact form as:\footnote{To gain some intuition, this comes from the fact that $cov(y_{i},\bar{y})=cov(y_{i},1/N\sum y_{j})=1/N\cdot(var(y_{i})+\sum_{j\neq1}cov(y_{i},y_{j})$), which in vector notation is equivalent to the row-sum of $\Sigma$, i.e. $\Sigma\mathbf{1}/N$. Furthermore, the variance of $\bar{y}$ is given by: $\mathbf{1}'\Sigma\mathbf{1}/N^2$. Taken together, we see that $cov(y_{i},\bar{y})/var(\bar{y})=N\Sigma\mathbf{1}/\mathbf{1}'\Sigma\mathbf{1}$}

\begin{equation}
\beta(\bar{y})=N\frac{\Sigma\mathbf{1}}{\mathbf{1}'\Sigma\mathbf{1}}
\end{equation}

As shown by \citet{Miranda1991}, the $\beta_i$ parameter plays an important role to assess the variance-reduction effects of index insurance for field $i$. A drawback of the $\beta_i$ parameter is that it depends on the field's own variance and hence is unbounded.\footnote{$\beta_i$ can be written as $\beta_i=\rho\sigma_i/\sigma_f$, where $\sigma_f$ denotes the variance of the index, and $\rho$ is the correlation coefficient between field's $i$ yields and the index.} This makes it challenging to build a basis risk metric based on $\beta_i$, and complicates further a comparison between fields or between insurance zones. Following \citet{ElabedBellemareEtAl2013}, we define rather basis risk as $BR_i(f)=1-R_i^2(y_i, f)$, the proportion of variance of field $i$ not explained by the index. Conversely, $R_i^2$ represents the insurable \emph{covariant risk}, i.e. the proportion of risk explained and insurable by the index. The $R_i^2$ is bounded between 0 and 1 and thus can be compared across fields. Denoting by $\mathcal{R}^{2}$ a $N \times N$ matrix whose diagonal elements are the individual $R_{i}^{2}$, the $R^{2}_i$ using the zone average as index, $f=\bar{y}$, can be written compactly  as:\footnote{The off-diagonal elements of $\mathcal{R}^{2}$ have no direct interpretation and are not used in our analysis. }

\begin{equation}
\mathcal{R}^{2}(\bar{y})=D^{-1/2}\Sigma1\left(1'\Sigma1\right)^{-1}1'\Sigma D^{-1/2}\label{eq:r2_output_mean}
\end{equation}

with $D$ a diagonal matrix containing the $\sigma_{i}^{2}$ from $\Sigma$.  While the above expressions are for the population $\beta$ and $\mathcal{R}^{2}$, their sample counterparts $\hat{\beta}$ and $\hat{\mathcal{R}}^{2}$ can simply be obtained by replacing $\Sigma$ with its sample counterpart, $S$.   This provides a compact way to obtain the $\hat{\mathcal{R}}^{2}$, and is numerically equivalent to obtaining the individual $\hat{\beta}_i$ and $\hat{R}_{i}^{2}$ by running $N$ regressions separately.

As noted above, the basis risk decomposition holds at the zone level, not the individual level. Hence, we need to choose an aggregation function for the zone-specific basis risk $BR$.  We take a simple average as the aggregation function, $BR=1/N\sum BR_i=1/N \sum (1-R^2_i)$, and define the average covariant risk $\overline{R^{2}}$ as the average $R_{i}^{2}$ over the fields within the same insurance zone:

\begin{equation}
\overline{R^{2}(\bar{y})}\equiv\frac{1}{N}\sum_{i}R_{i}^{2}=\Trace\left(\mathcal{R}^{2}(\bar{y})\right)/N\label{eq:r2_unweighted_mean}
\end{equation}

We can now ask which is the index $f$ maximizing the covariant risk $\overline{R^{2}(f)}$. Several studies have used the basis risk decomposition, assuming either that the optimal index was an input-based one \citep{ConradtFingerEtAl2015, YuVandeveerEtAl2019} or an output-based one such as the zone-average yield \citep{FlatnesCarterEtAl2018}. Intuitively, it is clear that an output-based index should lead to a lower basis risk compared to an input-based index, given that the output-based index relies on a richer set of information encompassing both weather components and yield shocks. Does this mean that the zone-average index $f=\bar{y}$ is the optimal index, that is the index leading to the highest $\overline{R^{2}(f)}$?

\begin{center}
\begin{figure}

\centering{}
\includegraphics[width=0.9\columnwidth]{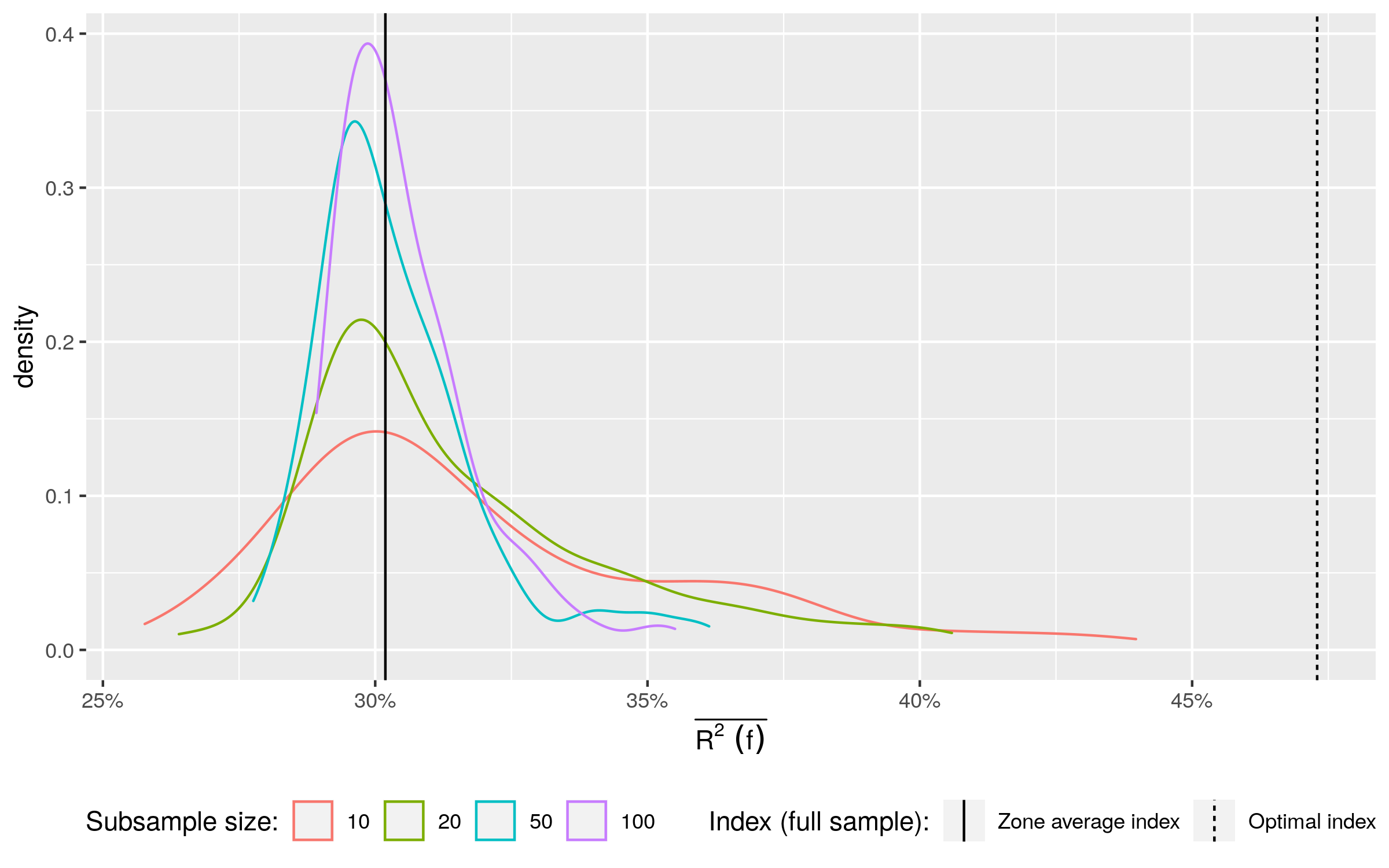}
\caption{Comparing $\overline{R^{2}(f)}$ using indices based on full and subsampled averages. \textmd{The solid and dashed lines represent the $\overline{R^{2}(f)}$ evaluated a respectively the zone-average and the optimal index. The density represent the $\overline{R^{2}(f)}$ using subsampled means as index, based on 200 replications for each subsample size. }}
\label{fig:simul-r2-submean}
\end{figure}
\end{center}

To verify whether the zone average is indeed the optimal index, Figure~\ref{fig:simul-r2-submean} shows a small experiment based on the yield data introduced below. In a nutshell, we consider an insurance zone with 200 fields, and measure the $\overline{R^{2}(f)}$ for various indices. We consider as index the zone-average  $f=\bar{y}$, as well as ``improper'' averages $f=\tilde{\bar{y}}$ based on random subsamples of size 10, 20, 50 and 100. We draw multiple random samples for each subsample size and show their density in Figure~\ref{fig:simul-r2-submean}. Contrary to intuition, the $\overline{R^{2}(f)}$ based on the improper averages $f=\tilde{\bar{y}}$ are often found to be higher than the one  based on the full-sample average, $\overline{R^{2}(\bar{y})}$. About half of the indices based on averaging just 10 observations do better than the full-sample mean. This is an important concern for the validity of the basis risk decomposition: assuming that the zone average $\bar{y}$ is the optimal index can lead to an ill-posed negative design risk!

What is then the index maximizing $\overline{R^{2}(f)}$ if it is not the zone-average $\bar{y}$? To find this out, let us consider more general output-based indices $f=Yw$ using any weight $w_i$ for field $i$, instead of focusing on the zone-average $\bar{y}$ restricting the weights to $w_i=1/N$. Looking at vectors with different weights is motivated by the experiment above, which showed that using a mean based on a subsample (i.e., using a weight vector where some entries are set to 0) can perform better than the full-sample mean. $\mathcal{R}^{2}(\mysmalleq{f}{Yw})$, the matrix containing the $R^2$ based now on the generalized index $f=Yw$, can be written as:

\begin{equation}
\mathcal{R}^{2}(\mysmalleq{f}{Yw})=D^{-1/2}\Sigma w\left(w'\Sigma w\right)^{-1}w'\Sigma D^{-1/2}\label{eq:r2_output_gen}
\end{equation}

Our statistic of interest is now $\overline{R^{2}(\mysmalleq{f}{Yw})}$, and we search for the optimal $w^\star$ maximizing $\overline{R^{2}(\mysmalleq{f}{Yw})}$, i.e. minimizing the average basis risk $\overline{1-R^{2}(\mysmalleq{f}{Yw})}$. Proposition~\eqref{prop:-The-optimal-vector} derives this optimal $w^\star$:

%\begin{prop}
%\textup{\label{prop:-The-optimal-vector}$w^{*}\equiv\argmin_{w}1/N\cdot tr(\mathcal{R}^{2}(w))=D^{-1/2}eigenvector(C).$
%That is, the optimal weight vector is the (rescaled) first eigenvector of the correlation matrix $C=D^{-1/2}\Sigma D^{-1/2}$. The corresponding optimal index is given by $\boldsymbol{y}^{*}\equiv Yw^{*}=YD^{-1/2}eigenvector(C)$, which is the principal component of the rescaled matrix $\tilde{Y}$ matrix $\tilde{Y}=YD^{-1/2}$. }\emph{Proof: see (\ref{proof_opt_vec_R})}. 
%\end{prop}

\begin{prop}
\textup{\label{prop:-The-optimal-vector} Let $w^{*}\equiv\argmax_{w} \Trace\left(\mathcal{R}^{2}(\mysmalleq{f}{Yw})\right)/N$ be the optimal vector of weights maximizing $\overline{R^{2}(\mysmalleq{f}{Yw})}$, $f^{*}\equiv Yw^\star$ the corresponding optimal index and C the correlation matrix $C\equiv D^{-1/2}\Sigma D^{-1/2}$. Then:
\renewcommand{\labelenumi}{1.\arabic{enumi}}
\begin{enumerate}
    \item $w^{\star}=D^{-1/2}eigenvector(C).$ That is, the optimal weight vector is the first eigenvector of the correlation matrix $C$ rescaled by a function of $D$, the diagonal matrix of variances from $\Sigma$.
    \item $f^{\star}=YD^{-1/2}eigenvector(C)$. %$\boldsymbol{y}^{*}=YD^{-1/2}eigenvector(C)$. 
    %That is, the optimal index  is the principal component of the rescaled matrix  $\tilde{Y}=YD^{-1/2}$.
    \item $\overline{R^{2}(f^{\star})} = \lambda_{1}^{C}/\sum_{i}\lambda_{i}^{C}$, where $\lambda_i^C$ is the ith largest eigenvalue of $C$.
That is, the objective function $\overline{R^{2}(f)}$ evaluated at the optimal predictor $f^{\star}$ is the (relative share) of the largest eigenvalue of  the correlation matrix between fields . 
\end{enumerate}}\emph{Proof: see proposition \ref{proof_opt_vec_R} in the appendix}. 
\end{prop}

Proposition (\ref{prop:-The-optimal-vector}) derives the optimal index $f^\star$ and quantifies the extent of the irreducible \emph{zonal} risk. We now have a well-defined decomposition of basis risk, where the zonal risk is defined as $\overline{1-R^{2}(f^\star)}$, i.e. the lower bound on the $\overline{1-R^{2}(f)}$ that any index will ever be able to reach for a given zone. Beyond that point, there is no possible improvement in the $\overline{R^{2}(f)}$ (or reduction in $\overline{1-R^{2}(f)}$), unless one starts to re-define the zone itself. Accordingly, design risk for any index $f$ can now be defined as $\overline{R^{2}(f^\star)}-\overline{R^{2}(f)}$ and is now guaranteed to be positive. 
Furthermore, (1.3) shows that the value of the irreducible  zonal risk is easily computed, given the relationship $\overline{R^{2}(f^{*})}=\lambda_{1}^{C}/\sum_{i}\lambda_{i}^{C}$: all that is needed is to build the correlation matrix between fields and compute the share of its first eigenvalue.
\footnote{Computation of the share of the first eigenvalue can furthermore be done very quickly, noting that  $XX^{'}$ and $X^{'}X$ share the same first $k=\min{(N,T)}$ eigenvalues and the same trace. This means that computing the first eigenvalue of a very large dataset containing thousands or even millions of fields can be done in just milliseconds. We provide R routines to compute this efficiently at XXX (\texttt{removed to preserve the anonymity of the reviewing process}).}
To illustrate this, the optimal value  $\overline{R^{2}(f^\star)}$ is shown as a dashed black line in the previous example comparing the $\overline{R^{2}(f)}$ using the full- or sub-samples average, see Figure~\ref{fig:simul-r2-submean}. It is clearly visible that while many indices have a higher $\overline{R^{2}}$ than the zone-average $\overline{R^{2}(\bar{y})}$, they all lie below the optimal index $\overline{R^{2}(f^\star)}$.

%Having derived this property, we realized that this property had been already established long time ago. \citet{MeredithMillsap1985} and \citet{JongKotz1999} for example discussed it, arguing that this property provides an alternative interpretation of principal component analysis (PCA) in term of providing the best common regressor in a system of regressions. The idea of finding a common variable to explain multiple variables might sound of little use in everyday statistical analyses, possibly explaining why this property has remained more of a curiosity. In the context of index insurance, however, this is exactly what we are after. 

Our result can be extended to an alternative objective function for basis risk, following the argument in  \citet{MeredithMillsap1985} and \citet{JongKotz1999}. Let us assume now that one wants to minimize the total sum of squared residuals  $TSSR(f)\equiv \sum_i SSR_i(f)$, instead of maximizing the average $\overline{R^2}$. It is easily seen that minimizing the total sum of squared residuals is equivalent to maximizing what we call the \emph{total} $R^{2}$, $\overline{\overline{R^{2}}}\equiv 1- \sum_{i}SSR_{i}/\sum SST_{i}$, where $SST_i$ is the sum of squares total for field $i$.\footnote{The sum of squares total is simply the quantity $\sum_i (y_i-\bar{y})^2$.} One can show (see \citealp{JongKotz1999}):

\begin{prop}
\label{prop:-eigen-cov-SSR} Let $w^{\star\star}\equiv\argmin_{w} \sum_i SSR_i(w) =  \argmax_{w}\overline{\overline{R^{2}(w)}}$. Then:
\renewcommand{\labelenumi}{2.\arabic{enumi}}
\begin{enumerate}
    \item $w^{\star\star}=eigenvector(\Sigma).$ The optimal weight vector is the first eigenvector of the covariance matrix $\Sigma$.
    \item $f^{\star\star}=Y\cdot eigenvector(\Sigma)$. The optimal index is the first principal component of  $\Sigma$.
    \item $\overline{\overline{R^{2}(f^{\star\star})}}=\lambda_{1}^\Sigma/\sum_{i}\lambda_{i}^\Sigma$
    \end{enumerate}
\end{prop}

Proposition~\eqref{prop:-eigen-cov-SSR} leads to very similar conclusions to Proposition \eqref{prop:-The-optimal-vector}. In particular, (2.3) shows that the covariant risk as defined by  $\overline{\overline{R^{2}}}$ can never be higher than the (share of the) first eigenvalue of the covariance matrix $\Sigma$. The total $\overline{\overline{R^{2}}}$ is closely related to the average $\overline{R^{2}}$, noting that the former is a variance-weighted average of the individual $R^{2}_i$, while the latter uses equal weights.\footnote{ One can write $\overline{\overline{R^{2}}}=\sum_{i}\alpha_{i}R_{i}^{2}$, with $\alpha_{i}=SST_{i}/\sum_{j}SST_{j}=var(y_{i})/\sum var(y_{j})$.} In practice, we found that the two criteria were giving very similar answers, and focus henceforth on the $\overline{R^{2}}$.

Theorems \eqref{prop:-The-optimal-vector} and \eqref{prop:-eigen-cov-SSR} make two similar contributions: (1) they each  derive the optimal index according to a specific criterion, and (2) they each indicate how to measure zonal risk as a simple function of eigenvalues. It is important to note that we view the second contribution as the most important, and do not necessarily advocate using the optimal index in practice. 
%First of all, note that the optimal vectors are not necessarily weighting the data with a convex combination, so that weights might be positive or negative and do not sum to one (in fact their squared values sum to 1). 
%Intuitively, it is not very clear how to interpret a negative weight.
Indeed, attributing distinct weights to every field defeats the purpose itself of index insurance, which is to provide a simple and cost-effective way to cover risk, without the need to measure yields on every field. Furthermore, the weights of the optimal indices do not necessarily sum to one and can even be negative. This implies that some fields might be disadvantaged when the optimal index is used. On the other hand, we view the second result showing how to compute the zonal risk as the main contribution of theorems \eqref{prop:-The-optimal-vector} and \eqref{prop:-eigen-cov-SSR}. Even if one does not intend to use the optimal index, having a clear and computationally simple indication of how the best index would perform can be very informative, indicating how far the elusive quest of the best index can go. This sets the basis to shift the focus from index quality to zone quality, which has been largely ignored in the literature. Moreover, the result formalizes the basis risk decomposition intuited by \citet{ElabedBellemareEtAl2013}.  As we illustrated above, absent this theoretical contribution, nonsensical situations can arise where ``imperfect'' indices perform better than the supposedly ``perfect'' one, leading to an ill-defined negative design risk.  Finally, the two theorems reveal a connection between index insurance and principal component analysis that was, to the best of our knowledge, not known. This allows to rephrase the premise of index insurance in statistical terms: in essence, one is assuming that the data can be adequately represented by a one-factor model. This is, in fact, the implicit assumption between most of the models on index insurance \citep{Miranda1991,Mahul1999,ElabedBellemareEtAl2013,BucheliDalhausEtAl2020}. Establishing this link opens the door to enrich further our understanding of index insurance by drawing on the rich literature on  factor models \citep{Bai2003,BaiLi2012}.

\section{Data\label{sec:Data}}

The main dataset we use consists of satellite-based maize yield estimates from 2016 to 2019 in Kenya. The data span all regions in Kenya that are cropped with maize. Yields were predicted with the SCYM model, which uses a crop model to estimate the relationship between satellite measures of greenness and final crop yield \citep{LobellEtAl2015,BurkeLobell2017,JinAzzariEtAl2017,JinAzzariEtAl2019}. The estimated relationship is then used to predict yields throughout the country using the satellite observations as input. When assessed against ground-truth measures in Kenya, \citet{BurkeLobell2017} report a prediction $R^{2}$ between .15 and 0.4,  \citet{JinAzzariEtAl2017SmallholderYieldHeterogeneity} report an $R^{2}$ of 0.4 against average yields at the county level and 0.3 against yields at the sub-county level. In a third iteration of the model, \citet{JinAzzariEtAl2019} report an accuracy of 0.5 against average yields at the county level. While there is room for progress, these numbers are encouraging, as predicting yields in smallholder systems with high heterogeneity of practices and very small field size is particularly challenging, and errors in ground-truth will lead to underestimation of satellite performance \citep{LobellDiTommasoEtAl2021}. Here we use estimates from a fourth iteration of the model, which were provided by Atlas AI. Figure~\ref{fig:map-yields-glocal}  shows the average yields over the area cropped with maize, as well as annual yields for a representative location. The SCYM model captures both large regional discrepancies as well as very localized differences in yields over the years. One potential concern with using satellite-based yield estimates is that measurement errors could influence results. For this reason, we also utilize crop-cut yields available for a subset of our study region to evaluate measurement error and demonstrate that our main conclusions are robust to plausible estimates of measurement error (see~\ref{subsec:robust_measure_error}).

Given the very high number of pixels covering the whole area, we opt for a field-level analysis by randomly selecting 200 pseudo fields in each of Kenya's lowest administrative divisions, the \emph{ward}. To define a pseudo-field, we consider only ``interior'' pixels whose adjacent pixels are also classified as maize, and randomly select 200 interior pixels from the pool of all interior pixels in a ward. This results in a dataset of $159\,600$ units in 798 wards observed over four years. 

\begin{center}
\begin{figure}
\centering{}\includegraphics[width=0.9\columnwidth]{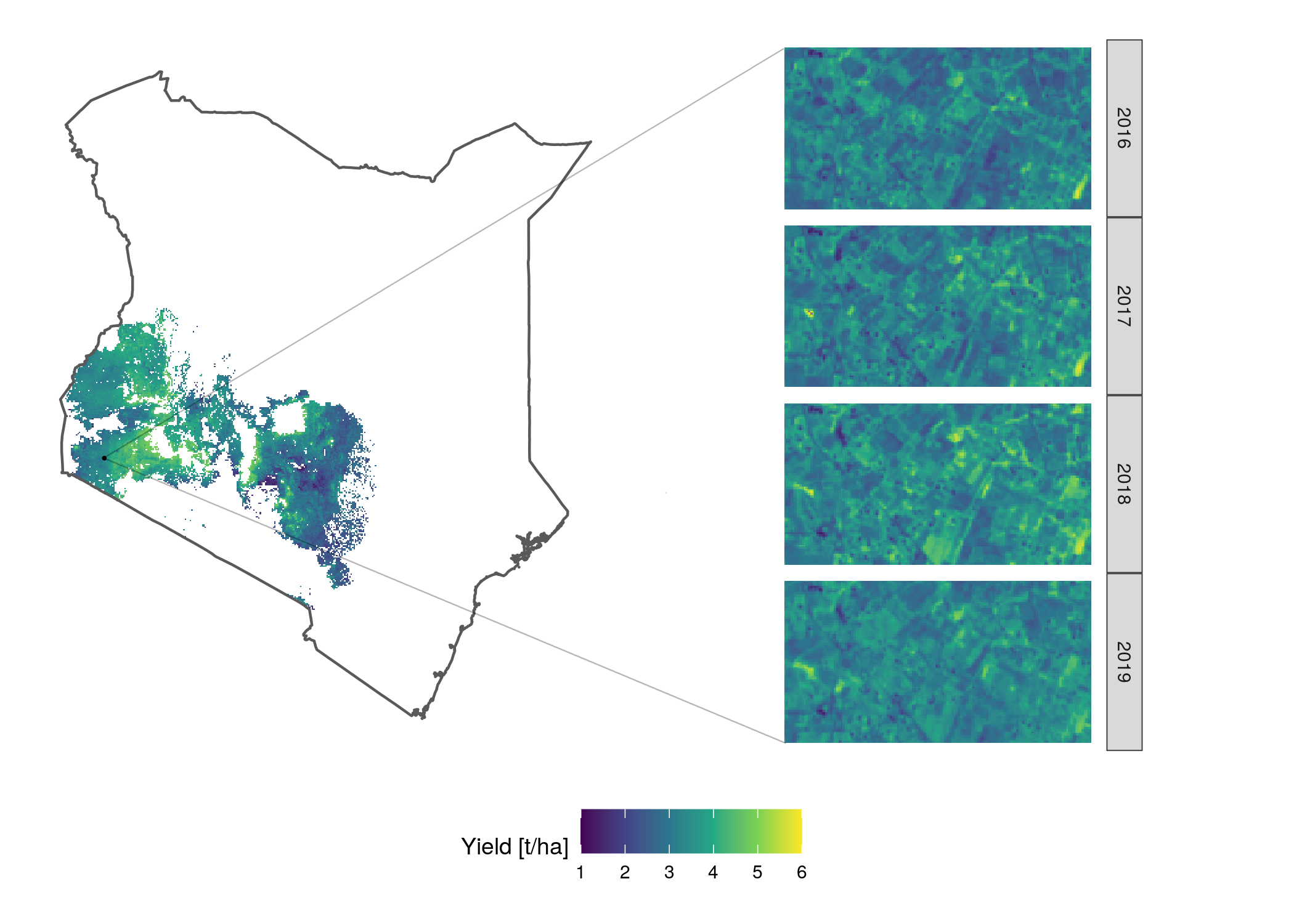}

\caption{Yields in Kenya as predicted by the SCYM model\label{fig:map-yields-glocal}.
\textmd{The left panel shows the SCYM yield predictions for the whole country, averaged over the four years. The right panel shows annual yields in a randomly selected region.}} %\david{this figure needs a scale bar on right. it'd also be nice to indicate on left where the sub-region is}
\end{figure}
\par\end{center}
%\begin{center}
%\begin{figure}
%\caption{Detailed by year, example\label{fig:map-yields-glocal}}

%\centering{}\includegraphics[width=0.9\columnwidth]{figures/yld_SCYM_local_rLoHySc}
%\end{figure}
%\par\end{center}

To investigate the impact of using smaller insurance zones, we make use of the very detailed administrative zones in Kenya.  Kenya is divided into three administrative levels, the county, the sub-county and the ward, which we will abbreviate as L1, L2 and L3. We sampled 200 fields within each of the 798 wards where we have available data, to yield a sample of $159\,600$ fields. These 798 wards are part of 182 sub-counties (L2) and 32 counties~(L1). Figure~\ref{fig:KEN-admin-4L} shows the administrative boundaries for each level. The sub-county (L2) and wards (L3) are relatively small administrative units, with an average area of 97 $km^2$ for the wards and of 427 $km^2$ for the sub-counties (see Table~\ref{tab:Zonal-risk-measures}). 

\begin{center}
\begin{figure}

\centering{}\includegraphics[width=0.9\columnwidth]{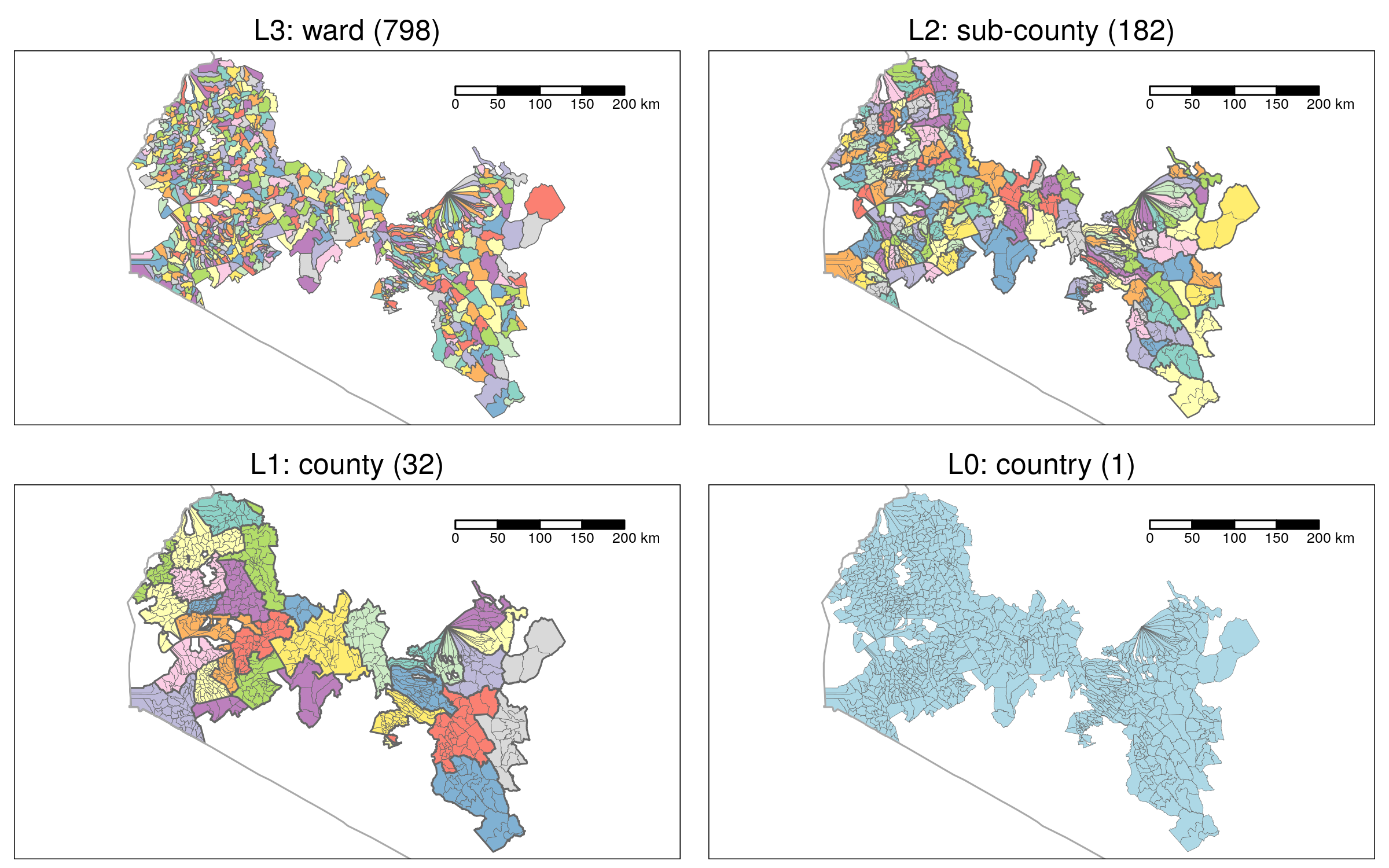}
\caption{Kenya administrative units\label{fig:KEN-admin-4L}.
\textmd{The numbers in parentheses show the total number of units for each administrative level.}}
\end{figure}
\par\end{center}

We also construct several weather-based indices to analyze design risk. These indices include minimum and maximum temperature (\emph{tmmn} and \emph{tmmx}), precipitation (\emph{precip}), and indicators of soil moisture (vapour pressure deficit, \emph{vpd} and soil moisture \emph{ssm}). We also include two popular vegetation indices, the normalized difference index (NDVI) and the related enhanced vegetation index (EVI). Strictly speaking, these last two indices are targeting the yields directly so they could be considered as output-based indices in our classification. However, given that they are only proxies, we consider those as input-based indices here. The data comes from four different sources: 1) the vegetation indices come from the MODIS satellite,
2) temperature, vapour pressure deficit  and precipitation are extracted from  TerraClimate \citep{AbatzoglouDobrowskiEtAl2018} and 
3) an alternative precipitation variable comes from the CHIRPS dataset \citep{FunkPetersonEtAl2015}. 
4) soil moisture is derived from SMAP \citep{ONeillChanEtAl2016}.

\section{Results\label{sec:Results}}

%We show here the results from the basis risk decomposition. We start by presenting estimates of the zonal risk in Section~\ref{subsec:Zonal-risk}, and then move to design risk in \ref{subsec:Design-risk}. 

\subsection{Zonal risk\label{subsec:Zonal-risk}}

To start, we present our estimates of the irreducible zonal risk $\overline{1-R^{2}(f^{*})}$, that is the lowest basis risk \emph{any} index can reach in a given zone. For ease of exposition, we focus on the complement of zonal risk, the covariant risk $\overline{R^{2}(f^{*})}$.   Recall that this is given by the relative share of the first eigenvalue of the correlation matrix for the first metric $\overline{R^{2}(f^{*})}=\lambda_{1}^{C}/\sum_{i}\lambda_{i}^{C}$, and by that of the covariance matrix for the second metric $\overline{\overline{R^{2}(f^{*})}}=\lambda_{1}^{\Sigma}/\sum_{i}\lambda_{i}^{\Sigma}$. In practice we found that the two metrics were very close and consequently report only the values for $\overline{R^{2}(f^{*})}$. 

Table~\ref{tab:Zonal-risk-measures} shows the covariant risk for various definitions of the insurance zone. We start by considering a national insurance zone, i.e. use a single index for the whole country. We find a $\overline{R^{2}(f^{*})}$ of 41.1\%, implying an irreducible zonal risk $\overline{1-R^{2}(f^{*})}$ of 58.9\%. Finding that barely 40\% of the total variation in yields is explained by the index seems rather small, but this could be due to the use of a very large insurance zone covering heterogeneous regions. To gain more insights into this heterogeneity, we vary the definition of the insurance zone using either the county, sub-county or ward level. To do so, we compute the $\overline{R^{2}(f^{*})}$ for each administrative unit and take their average over the whole country. Looking again at Table~\ref{tab:Zonal-risk-measures}, we see that using instead county (L1) boundaries for the insurance zones, the $\overline{R^{2}(f^{*})}$ increases from 41\% to 48\%, reducing the zonal risk from 59\% to 52\%. Using sub-county (L2) or wards (L3) as insurance zones further reduces  the zonal risk, yet to a smaller extent. At the finest level (L3 wards), the zonal risk averaged over the 798 wards units is still 48\%, meaning that the best local index is only capturing half of the local variations in yields. To put this number in context, \citet{StiglerLobell_cropUS_wp} report that at the lowest administrative level in the United States---the county---the average zonal risk is close to 25\% for maize. Noting that a US county is about 17 times bigger on average than a ward in Kenya, this comparison suggests a substantial local heterogeneity in Kenya. 
%% DATA REF: see 4_3_zonalRisk_byL_compareUS_rJuWwUs

\begin{table}
\caption{Zonal risk measures\label{tab:Zonal-risk-measures}}

\centering

\begin{tabular}{llrrrll}
  \toprule
Level & Id & N units & N fields (ave) & Area $[km^2]$ & $\overline{R^2(f^\star)}$ & Zonal risk $\overline{1-R^2(f^\star)}$ \\ 
  \midrule
Country & L0 & 1 & 159600 & 77739 & 41.1\% & 58.9\% \\ 
  County & L1 & 32 & 4988 & 2429 & 47.9\% & 52.1\% \\ 
  Sub-county & L2 & 182 & 877 & 427 & 49.7\% & 50.3\% \\ 
  Ward & L3 & 798 & 200 & 97 & 52.2\% & 47.8\% \\ 
   \bottomrule
\end{tabular}

\end{table}

The analysis so far used administrative boundaries to delineate insurance zones. However, administrative boundaries tend to be arbitrary, and do not necessarily reflect natural homogeneous zones in terms of agronomic conditions. This is particularly true in the Eastern part of our sample, where wards are in a ``wheel and spoke'' configuration, where the center is Mount Kenya. Given the high density of fields we have in the sample, we can investigate the effect of using alternate definitions of the zones, as well as using even smaller zones. 

To create smaller insurance zones, we draw a circle around each field and vary the size of the circle's radius from 50 meters to 50000m. Using these zone definitions is perhaps an unfeasible scenario but nonetheless provides a useful benchmark.\footnote{An alternative benchmark could use small non-overlapping zones instead of using a field-specific zone as we do here. Segmenting the whole maize area into contiguous homogeneous zones is, however, a very difficult problem computationally that we leave for future research. } To ensure that there are enough fields in each zone, we extend our definition of pseudo fields to encompass all maize pixels within a given circle instead of only considering the 200 pseudo fields initially sampled. We then compute our measure of zonal risk among all pseudo fields found within each circle. To ensure that we do not include pixels that belong to the central field, we repeat the exercise computing the same measure yet excluding pseudo fields less than 50m from the reference field.

Figure~\ref{fig:Average-R2-with_radius} shows the covariant risk averaged among all fields for various radius sizes. Lines in black represent the average ``radius'' of the administrative zones\footnote{We use as measure of a polygone's average ``radius'' the average length of the segments linking each edge to the centroid of the polygon.} together with the average zonal risk within the administrative units. For zones with a radius smaller than 1000m, the global $\overline{R^2}$ is substantially higher than values observed using administrative boundaries. When the circle's radius approaches the average radius of the administrative zones, results become very  similar. For small zones, excluding the 50m inner circle from the computation leads to lower covariant risk. However, the difference vanishes quickly and becomes imperceptible with radius sizes of 1 km or more.  

%To do so, we now consider zones formed from the K nearest-neighbors (KNN) sampled fields, where we vary the number $K$ of neighbors from 200 to 20. Using 20 fields is perhaps an unfeasible scenario, as it considers very small neighborhoods that are field-specific, but nonetheless provides a useful benchmark. Figure~\ref{fig:Average-R2-with_radius} shows a boxplot of the $\overline{R^{2}(f^{*})}$ for each zone definition, reproducing also the values using the ward (L3) as comparison.\footnote{For the sake of comparison, the units of observation in the boxplot are ward aggregates of the field-specific KNN zones. } Starting from the comparison between zones with 200 fields using the administrative L3 boundaries versus using 200 KNN, the average improvement is relatively small (from 52.2\% to 53.2\%). However, larger improvements are observed as zones become smaller, with the $\overline{R^{2}(f^{*})}$ for the smallest zones with 20 neighbors averaging 60.9\%, implying an average basis risk just below 40\%. While this is a sizable reduction of 10\% in zonal basis risk, one should keep in mind that it is obtained under very favorable conditions using very small zones that are furthermore specific to each field. In that sense, obtaining only a zonal homogeneity of 60\% suggests the presence of strong heterogeneity even at a very local level. 

\begin{center}
\begin{figure}

%viewport=0bp 0bp 2400bp 1420bp,clip,
\centering{}
\includegraphics[trim=0 0 0 12, clip, width=0.9\columnwidth]{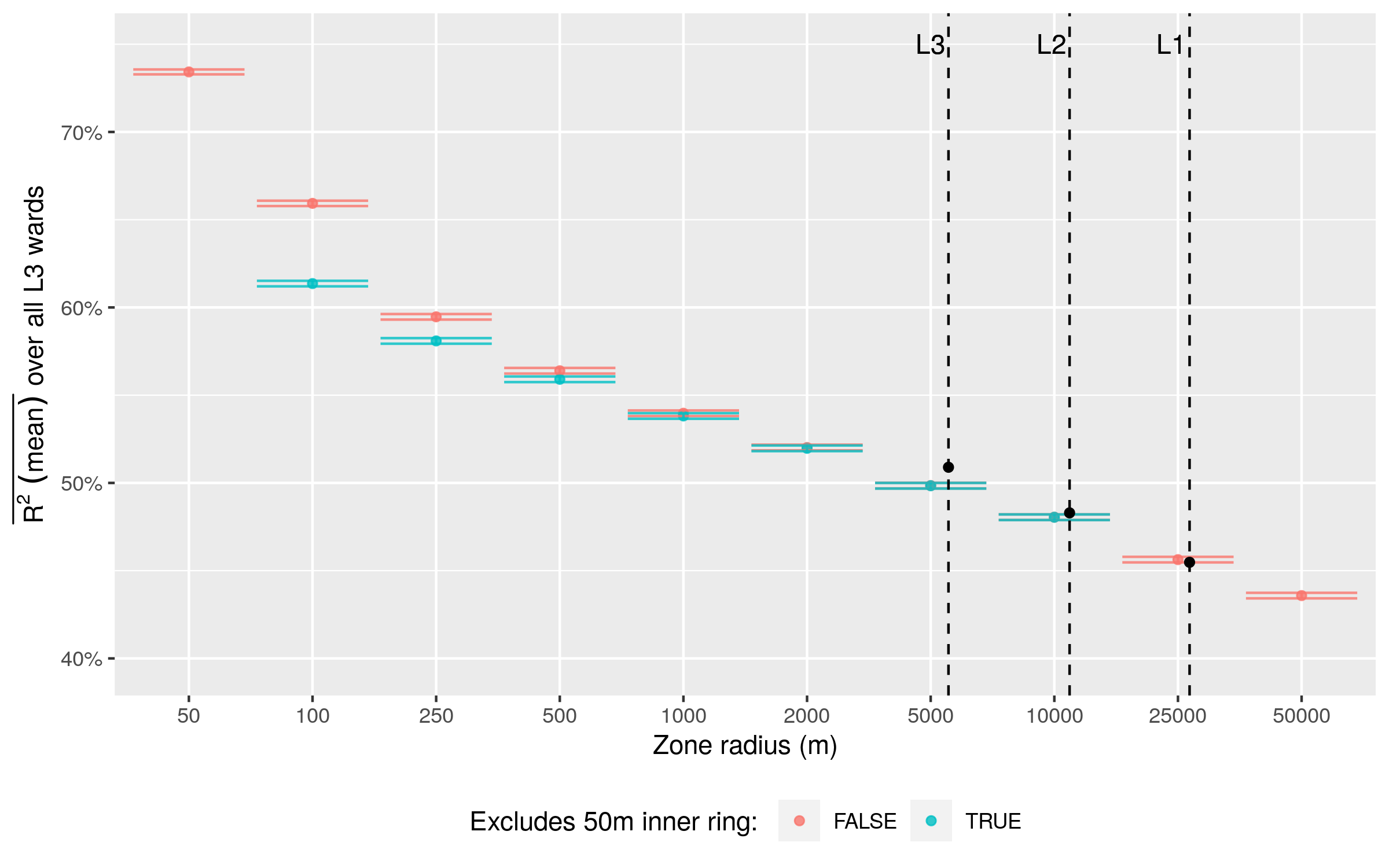}
\caption{Average $R^{2}$ using local zones.\label{fig:Average-R2-with_radius}
\textmd{The dashed lines  represent the average ``radius'' of the administrative levels. The black dots represent the corresponding average zonal $R^{2}$ at each administrative level.}}
\end{figure}
\par\end{center}

The analysis so far focused on the mean covariant risk averaged over all insurance zones.  However, the average might conceal important variations between the insurance zones. A map of zonal risk in each ward (Figure~\ref{fig:Map-Zonal-risk}) reveals a large variation, with values of $\overline{R^{2}(f^{*})}$ varying from 35\% to 85\%. The map also reveals  important spatial patterns, indicating that the covariant risk $\overline{R^{2}(f^{*})}$ is higher in the central and eastern part than in the western region. In general, zonal homogeneity seems to be clustered within regions, with clusters of high as well as very low values in the western part. 

\begin{center}
\begin{figure}

%viewport=0bp 0bp 2400bp 1350bp,
\centering{}
\includegraphics[width=0.9\columnwidth]{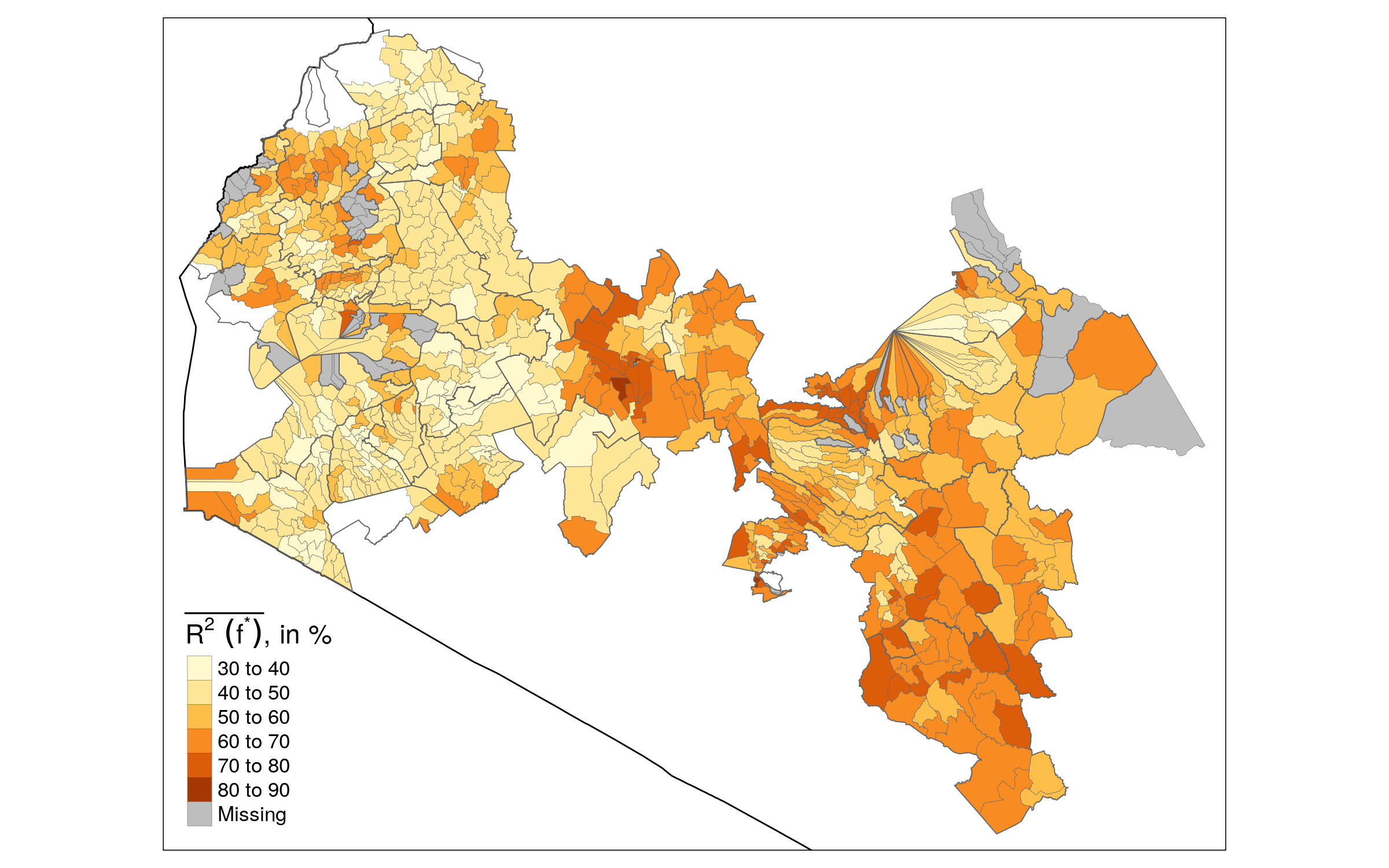}
\caption{Covariant risk $\overline{R^{2}(f^{*})}$ using wards (L3) administrative boundaries}
\label{fig:Map-Zonal-risk}
\end{figure}
\par\end{center}

%Overall, we find that the irreducible zonal basis risk is rather high in Kenya. Using the finest administrative units as insurance zone, we show that the average irreducible basis risk is about 48\%. Using even finer zones based on circles surrounding each field, we fin

\subsection{Design risk\label{subsec:Design-risk}}

We investigate now the extent of design risk, that is the additional risk introduced by using a feasible index instead of the optimal one. The feasible index can be either output- or input-based. For output-based indices, we consider the standard zone-average yield (also called area-yield), a scheme  implemented in the USA insurance program, and which \citet{WB_Clarke_Kenya_2015} suggested implementing in Kenya. For input-based indices, we consider various indices including weather indices (temperature, precipitation, etc) and satellite vegetation indices. 

Starting with an index based on the average yield, Figure~\ref{fig:design_risk_r2_mean_opt_SCYM_l3} compares the basis risk measures using either the optimal index $\overline{R^{2}(f^{*})}$ or average yields $\overline{R^{2}(\bar{y})}$ for each of the 798 wards. As predicted by the theory developed above, the $\overline{R^{2}}$ based on the optimal index is always greater than or equal to the $\overline{R^{2}}$ based on the zone-average index. Interestingly, for high values of the underlying optimal index, the two measures are very close. For lower values of the optimal index, $\overline{R^{2}(\bar{y})}$ can be substantially lower than the $\overline{R^{2}(f^{*})}$. This result implies that in zones that already have a high zonal risk (a low covariant risk), using a zone-average index adds a supplementary amount of design risk, thus increasing total basis risk. For example, in some wards with a zonal basis risk of 60\%, the total basis risk goes up to 70\% using an average-yield index rather than the optimal index. Conversely, the potentially insurable covariant risk, initially of 40\%, would be reduced to an actually insured covariant risk of 30\%. 

\begin{center}
\begin{figure}
%viewport=0bp 0bp 2400bp 1425bp,clip,
%\centering{}
\includegraphics[trim=0 0 0 18, clip, width=0.9\columnwidth]{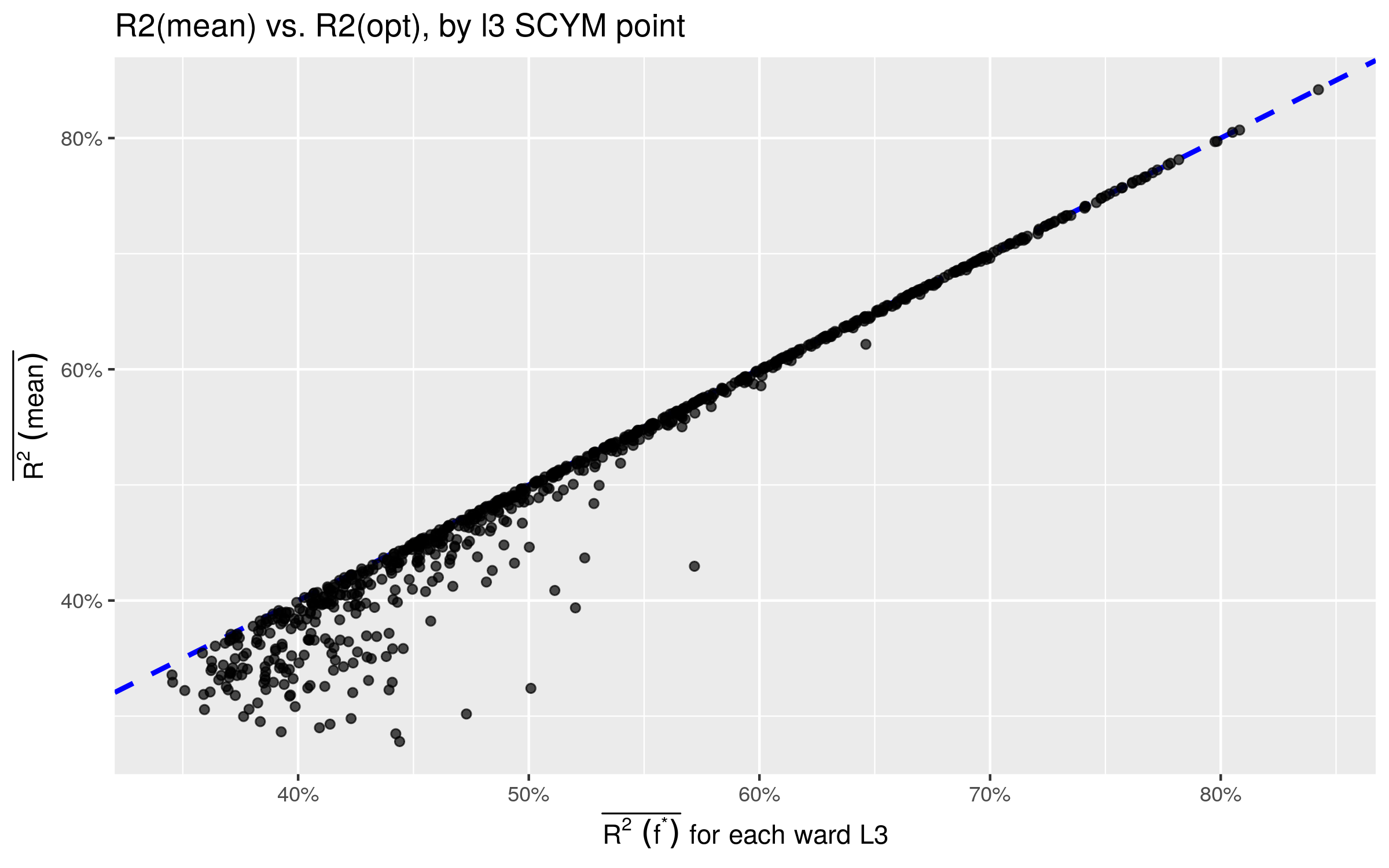}
\caption{Design risk: Comparing $\overline{R^{2}}$ from optimal and zone-average index}
\label{fig:design_risk_r2_mean_opt_SCYM_l3}
\end{figure}
\par\end{center}

Table~\ref{tab:Design-risk-weather-R2} shows the average $\overline{R^{2}(f)}$ measures at the ward level (L3) for various indices $f$ together with the $\overline{R^{2}(\bar{y})}$ obtained using the zone-average $f=\bar{y}$ and the optimal index $f^{*}=\mathbf{Y}w^{*}$. The column \emph{N wards} indicates the number of wards for which we have full data. The full data is not always available due to missing observations in the weather products. Turning to the main column $\overline{R^{2}(f)}$, the two first rows under the \emph{Yields} category report the numbers reported above, of $\overline{R^{2}(f^\star)}=52\%$ for the optimal index and $\overline{R^{2}(\bar{y})}=51\%$ for the index using zone averages. We move now to input-based indices, which include various simple measures of weather (precipitation, min and max temperature), soil moisture, vapour pressure deficit (VPD) or vegetation conditions (EVI and NDVI). The indices are built by using a growing-season temporal average for each pseudo field and are then spatially averaged over the pseudo fields within each ward.

We observe a very sharp drop in the $\overline{R^{2}(f)}$ For the input-based indices, with all indices averaging around 11\%, This implies a total basis risk close to 90\%. With such a high basis risk for input-based indices, implementation of a scheme with a weather index would likely encounter limited success. Surprisingly, we see little differences in the indices themselves, all leading to an average $\overline{R^{2}(f)}$ around 11\%. Where we see some differences is in the correlation between the $\overline{R_{w}^{2}(f)}$ and the $\overline{R_{w}^{2}(f^{*})}$ at the ward level $w$, shown in the column \emph{Cor with opt index}. The temperature and precipitation indices have the lowest correlation with the index, while measures based on vegetation (NDVI, EVI) or soil moisture (SMAP) have higher correlations, close to 50\%. This suggests that the \emph{zone quality }(the $R^{2}$ from the optimal index) is weakly linked to the total basis risk (the $R^{2}$ from the optimal index). 

Several factors can contribute to the high estimated values for design  risk. First, we used relatively simple weather indices, based on average values over the growing-season months. Additional effort on index design would likely result in better indices. A second explanation could lie in the resolution of the data. While the resolution of the yield data is at $10m\times10m$ (from the Sentinel 2 satellite), the resolution of the other products ranges from $500m^{2}$ to $10km^{2}$. With broader resolutions, the probability that a pixel includes information from other fields as well as from non-agricultural elements (roads, forests, etc) inevitably increases. This ``pixel contamination effect'' could partly account for the high design risk observed here. A final explanation could come from measurement error in the SCYM yield data, which would attenuate the correlation between predicted yields and the weather indices, hence inflating design risk. While carefully disentangling each explanation lies beyond the scope of this paper, the next section addresses measurement error in the SCYM data.

%Our data shows that there is a strong local heterogeneity in yields, implying that within a $500m^{2}$ zone, there is a large part of the variance in yields that a single weather value cannot account for. However, noting that in both cases the final index is constructed by averaging all values within the zone, this explanation might not fully account for the low $R^2$ we encounter. 
%Understanding the effect of the pixel resolution and the scale of aggregation remains yet to be fully understood, and we leave this for future research.

%Noting that the indices here are built by averaging the weather pixels within each zone, the impact of averaging over $500m^{2}$ or $10km^{2}$ pixels will largely depend on the homogeneity of the variables themselves. In the case of temperatures, 

 %Understanding the effect of the pixel resolution and the scale of aggregation remains yet to be fully understood, and we leave this for future research.

\begin{table}

\caption{Design risk: $\overline{R^{2}(f)}$ from weather\label{tab:Design-risk-weather-R2}}

\begin{centering}
% latex table generated in R 4.1.1 by xtable 1.8-3 package
\begin{tabular}{lllrrr}
  \toprule
Dataset & Variable & Resolution & N wards & $\overline{R^2(f)}$ & Cor with opt index \\ 
  \midrule
Yields & Optimal Index & 10m & 798 & 52.2 &  \\ 
   & Zone mean & 10m & 798 & 50.9 & 0.98 \\ 
  MODIS & EVI & 500m & 733 & 11.7 & 0.55 \\ 
   & NDVI & 500m & 733 & 11.6 & 0.56 \\ 
  CHIRPS & Precip & $\sim$5000m & 798 & 10.9 & 0.50 \\ 
  TerraClimate & Precip & $\sim$4000m & 798 & 10.0 & 0.38 \\ 
   & Temp min & $\sim$4000m & 798 & 11.7 & 0.48 \\ 
   & Temp max & $\sim$4000m & 798 & 11.0 & 0.43 \\ 
   & VPD & $\sim$4000m & 798 & 11.9 & 0.47 \\ 
  SMAP & Surface soil moisture & $\sim$12000m & 784 & 11.1 & 0.57 \\ 
   & Subsurface soil moisture & $\sim$12000m & 784 & 11.1 & 0.56 \\ 
   \bottomrule
\end{tabular}

\par\end{centering}
\end{table}

\section{Robustness checks}\label{sec:robustness}

\subsection{Impact of measurement error}\label{subsec:robust_measure_error}

While using satellite data provides us with both a very detailed and very  broad dataset compared to alternative data sources, it has the potential drawback of being less precise than traditional ground truth measurements. The analysis so far was carried assuming that the satellite data was exempt of measurement error. To investigate the impact of measurement error, we leverage a dataset of crop cuts collected by the One Acre Fund Foundation (1AF). This is a particularly rich crop cuts dataset containing 14 000 samples collected over the same four years, yet it comes with its own limitations. The sampled area changed from year to year and geo-referencing of the fields is too imprecise to attribute a crop cut to a specific field. This leads us to construct a panel dataset aggregating crop cuts at the ward (L3) level instead of using the field as unit of analysis. Using only wards that have four years of observations and at least three crop cuts each year, we end up with a dataset of 107 units. The SCYM dataset is constructed in the same way, aggregating our pseudo fields at the ward level for the same 107 units. 

To assess the measurement error structure, we estimate the following regression:

\begin{equation}\label{eq:measure_error}
    y_{it}^{SCYM} = \alpha + \gamma y_{it}^{cCut}+\varepsilon_{it}
\end{equation}

We report two statistics from regression \eqref{eq:measure_error}, the correlation between the SCYM and the crop cuts $\rho$, as well as the slope coefficient $\gamma$. The slope parameter $\gamma$ is of particular interest, since it informs about the conditional bias of the predictions. A value close to 1 indicates that the predictions are ``conditionally unbiased''. On the other hand, the intercept parameter  $alpha$ is of less interest, since presence of an unconditional bias $alpha\neq 0$ does not impact estimates of variance, covariance or correlation. 
The first row of table~\ref{tab:measurment_error} shows that for the pooled regression \eqref{eq:measure_error}, the correlation between the predicted and ground-truth value is 0.39. This is somewhat smaller than the accuracy reported in the original papers \citep{JinAzzariEtAl2017SmallholderYieldHeterogeneity, JinAzzariEtAl2019}, which were based on data from 2015 to 2017. The slope coefficient is 1.2, and the hypothesis that it is different from 1 can be rejected at the 10\% level, yet not at 5\%. 

\begin{table}

\caption{SCYM measurement error: comparing SCYM against crop cuts from 1AF\label{tab:measurment_error}}

\begin{centering}
\begin{tabular}{lrlrr}
  \toprule
Regression Type & Cor & Coefficient & Gamma & $P(\gamma=1)$ \\ 
  \midrule
Pooled & 0.39 & $\gamma$ & 1.20 & 0.06 \\ 
  Temporal & 0.50 & $\overline{\gamma_i}$ & 1.05 & 0.77 \\ 
  Spatial & 0.37 & $\overline{\gamma_t}$ & 1.14 & 0.19 \\ 
   \bottomrule
\end{tabular}

\par\end{centering}
\end{table}

The slope coefficient in the measurement error regression \eqref{eq:measure_error} corresponds to  a pooled panel, and is hence a mixture of cross-sectional and temporal regressions \citep{Wooldridge2010}. As our basis risk measures are essentially based on regressions over time, we might rather be interested in $\overline{\rho_i}$ and $\overline{\gamma_i}$, the average coefficients from a regression over time, i.e. $y_{it}^{SCYM} = \alpha_i + \gamma_i y_{it}^{cCut}+\varepsilon_{it}$. We hence compute each regression separately, and report the averages $\overline{\rho_i}$ and $\overline{\gamma_i}$ in the second row of Table~\ref{tab:measurment_error}.\footnote{To do inference on $\overline{\gamma_i}$, we exploit the  fact that $\overline{\gamma_i}$ is closely related to the $\gamma^{FE_i}$ obtained using time fixed effects  (see \citealt{GibbonsCharlesCarlosEtAl2019}), and hence use a fixed-effect estimator to test $\overline{\gamma_i}=1$  } The average temporal correlation $\overline{\rho_i}$  is 0.5, higher than the pooled $\rho$ which was 0.39. The $\overline{\gamma_i}$ parameter is also closer to 1, and now statistically not distinguishable from 1. The same exercise can be repeated for the cross-sectional variation, tracking how SCYM correlates on average with the crop cuts over space. The corresponding parameter $\overline{\rho_t}$ and $\overline{\gamma_t}$ are shown in the third row. The average correlation with the ground truth data $\overline{\rho_t}$ is 0.37, closer to the 0.39 we found for the pooled correlation $\rho$. The slope parameter $\overline{\gamma_t}$ is not statistically different from 1 either. Taken together, these findings indicate that while still leaving room for progress, SCYM's accuracy is higher for  temporal than for cross-sectional variations, and has no conditional bias. These two features are particularly important in our context, where we focus on estimating basis risk from regressions over time.

The ground truth dataset can also be used to assess the sensitivity of our results to measurement error. To do so, we compute our main statistic of interest, $\overline{R^2(f^\star)}$, on both the ground truth and the corresponding SCYM dataset. Recall that both are constructed by considering a ward as a unit of analysis, and hence do not directly correspond to the values reported elsewhere in the paper.
We repeat the same exercise considering the sub-counties (L2) or counties (L1) as units of analysis instead. Table~\ref{tab:measurment_eigen} reports the values computed at the three levels of analysis. Comparing the $\overline{R^2(f^\star)}$ estimated on the crop cut or SCYM pseudo-field data, we notice that SCYM under-estimates the covariant risk in two cases (for the analysis at L2 and L1), and over-estimates it at the L3 level. In all these cases, we see that the bias is moderate and never higher than 0.1. Were this bias extrapolated to our analysis at the field level, it seems unlikely that our general finding of a low covariant risk would radically change without measurement error. As a last robustness check, we include an alternative version of the SCYM data, where the units are constructed by averaging all the pixels within a ward instead of restricting it to our 200 randomly sampled pseudo fields. Column \emph{SCYM: all pixels} shows that these values are relatively close to the ones obtained with the pseudo fields. 

\begin{table}

\caption{SCYM measurement error: covariant risk estimated on crop cuts versus SCYM data\label{tab:measurment_eigen}}
\begin{centering}
    \begin{tabular}{lrrrr}
  \toprule
 &&\multicolumn{3}{c}{$\overline{R^2(f^\star)}$} \\
\cmidrule(l){3-5}
Analysis level & N. units  & Crop Cut & SCYM: pseudo fields & SCYM: all pixels \\ 
  \midrule
L1 &  14 & 0.52 & 0.45 & 0.50 \\ 
  L2 &  49 & 0.52 & 0.48 & 0.46 \\ 
  L3 & 103 & 0.39 & 0.50 & 0.51 \\ 
   \bottomrule
\end{tabular}

\par\end{centering}
\end{table}

\subsection{Relation to other measures}

The linear correlation measures we use so far have limitations in the context of index insurance. Arguably, it is more important for an index to accurately predict yield losses than to predict good harvests. Various approaches have been suggested to take this into account, ranging from quantile regression \citep{ConradtFingerEtAl2015} to more sophisticated left-tail dependence indices \citep{Bokusheva2018}. Another approach seeks to directly assess the benefits of index insurance using an expected utility (EU) framework \citep{CarterChiu2018}. Under the standard assumption of risk-aversion and concave risk-utility function, an index that accurately predicts yield losses will be preferred to one that accurately predicts good harvests. While more theoretically consistent, deriving the optimal index according to a quantile loss function or an EU one is very challenging, and we leave this for future work. However, we can at least examine how our correlation-based measures compare with more refined ones. We opt for two measures: a quantile-based correlation measure similar to our linear correlation measure and an expected utility measure.  

Computing the quantile measures of basis risk and the EU measures of insurance benefits requires several adjustments and assumptions. For the quantile measure, the previous literature has mainly used quantile regression to determine the indemnity function \citep{ConradtFingerEtAl2015,BucheliDalhausEtAl2020}, yet we are not aware of previous work using quantile regression to simply measure basis risk. To do so, we derive a quantile version of our total $R^{2}$ measure based on the quantile pseudo $R^{2}$ developed by \citet{KoenkerMachado1999}. \citet{KoenkerMachado1999} suggest a pseudo $R^{2}(\tau)=1-V(f,\tau)/V(1,\tau)$ at quantile $\tau$, where $V(f,\tau)$ is the quantile analogue to the sum of squared residuals (SSR) in a linear regression with variable $f$. Likewise, $V(1,\tau)$ is the quantile analogue to the total sum of squares from a regression including only a constant. In a similar way to our average $\overline{R^{2}(f)}$, we define the average quantile pseudo $R^{2}$ as $\overline{R_{q}^{2}(f)}=1-\sum V_{i}(f,\tau)/V_{i}(1,\tau)$, and use the value of $\tau=0.3$ following previous literature \citep{BucheliDalhausEtAl2020}.

Unlike the correlation-based measures, computing the expected utility measures of index insurance requires multiple assumptions. One needs to define an indemnity function, assume a specific form for the utility function and decide which benchmark to use as comparison. For the indemnity function, we use the function in \citet{Miranda1991}, $\ensuremath{I_{t}=\max(\lambda\bar{y}_{\cdot\cdot}-\bar{y}_{\cdot t},0)}$, where $\bar{y}_{\cdot t}$ corresponds to the zone-average yield at time t, and $\bar{y}_{\cdot\cdot}$ to its value averaged over time. $\lambda$ is the trigger level, that is the threshold (in percentage) below which an indemnity is paid. We set this level at a relatively high value of 90\%, to ensure that payouts occur in our sample. We also use a fair premium, i.e. we set it equal to the expected indemnity, $\pi=E[I_{t}]$. Yields with insurance $\tilde{y}$ are then given by $\tilde{y}_{it}=y_{it}+I_{t}-\pi$. For the utility function, we follow previous literature \citep{WangHansonEtAl1998,DengBarnettEtAl2008,FlatnesCarterEtAl2018} and use a constant relative risk aversion (CRRA) iso-elastic utility function, with a parameter of 1.5. Using this function, we compute the expected utility of the post-insurance yields $\tilde{y}_{it}$, $E[u(\tilde{y}_{it})]$, and derive the corresponding certainty equivalent (CE).\footnote{The certainty equivalent of a lottery $y$ is the sure amount $c$ such that: $u(c)=E[U(y)]$ holds. It has the advantage of being expressed in the same unit as $y$, whereas EU measures are expressed in ``utility units''.} The last ingredient is the choice of the benchmark, i.e. whether we are comparing the CE of index insurance to the CE of yields without insurance or to the CE of yields with individual insurance. As showed by \citet{StiglerLobell_cropUS_wp}, a comparison of index insurance to no insurance will be particularly sensitive to the variance of yields within that zone. That is, a field with a high variance yet little correlation with the index might perform better in that metric than a field with high correlation yet low variance. Conversely, comparing index insurance to individual insurance amounts to comparing two different variance reduction tools. Therefore, it is less sensitive to the raw variance of a given field. Given that our $R^{2}$ measure of basis risk is variance-free, the second approach seems more natural in our context. We hence adopt the \emph{farm-equivalent risk coverage }metric proposed by \citet{StiglerLobell_cropUS_wp}, which seeks the highest level of individual insurance compatible with an index insurance at 90\% level.

A final issue with the post-insurance EU measures is that they critically depend  on whether an indemnity is paid. This in turn depends on whether the minimum value is above or below 90\%, that is, whether the lowest observed (zone-average) yield is below 90\% of its mean. If not, there is no insurance payment at all, making the evaluation of the benefit of insurance impossible. Unfortunately, with just four years of data, this happens relatively often: 66\% of the L3 wards never experience a minimum value below 90\% of their mean. To address this problem, we proceed to a data simulation. We first estimate regression (\ref{eq:Mir_reg}) relating field's i to the ward average, extracting the $\hat{\alpha}_{i}$, $\hat{\beta}_{i}$ and $\hat{\sigma}_{i}$. We then simulate 30 years of ward-average yields $\hat{y}_{jt}$, based on a normal distribution with mean and variance parameters estimated from the yield data. We finally simulate individual yields by plugging in the field-level estimates and the simulated means, that is $\hat{y}_{ijt}\sim\mathcal{N}\left(\hat{\alpha}_{i}+\hat{\beta}_{i}\hat{y}_{jt},\hat{\sigma}_{i}\right)$. As a result of this simulation, we now only have 30\% of wards for which the relative minimum is never below 90\%.

\begin{center}
\begin{figure}

\centering{}\includegraphics[width=0.9\columnwidth]{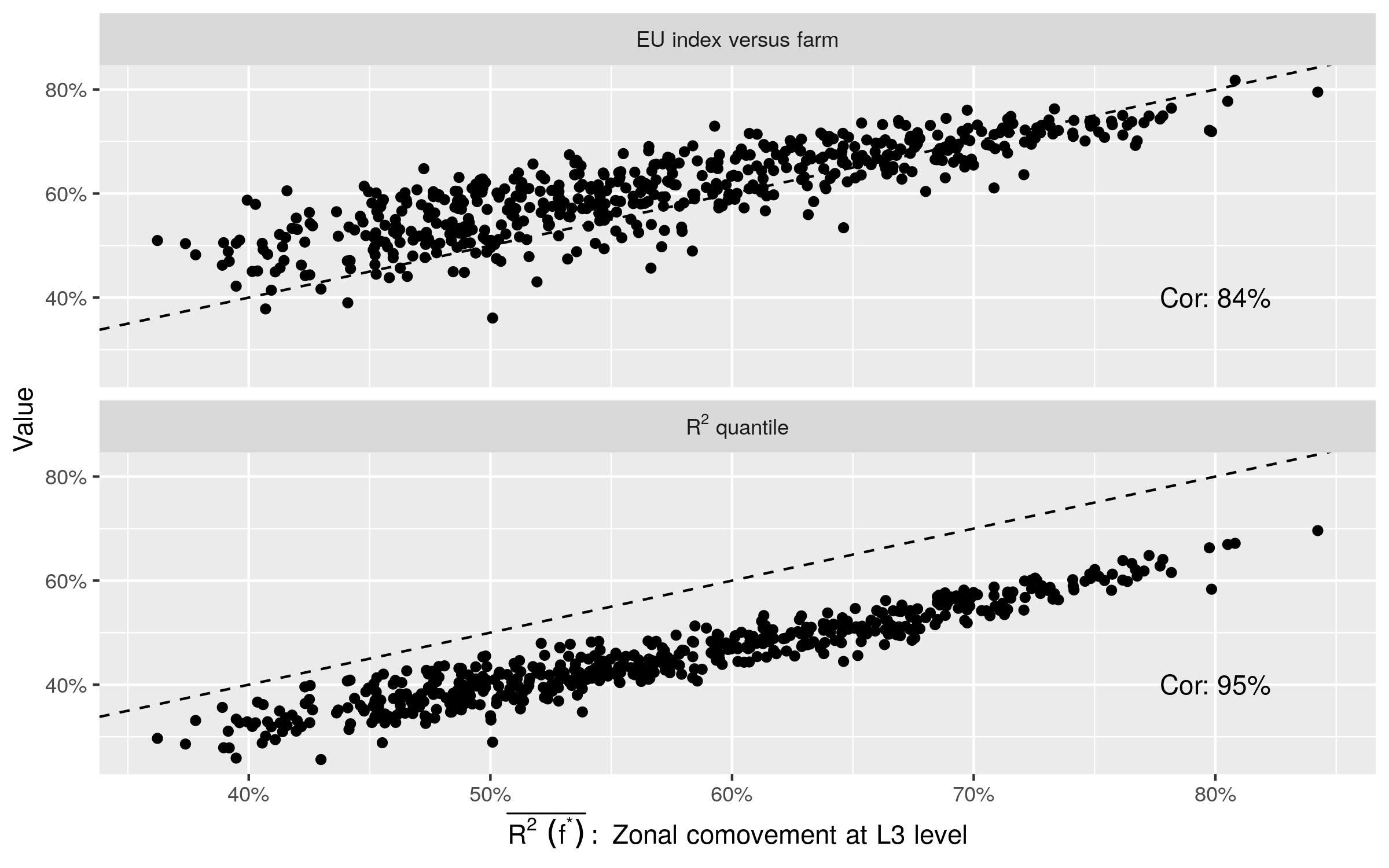}
\caption{Comparing $\overline{R^{2}}$ with quantile $\overline{R_{q}^{2}}$
and EU measures\label{fig:robust-EU-r2Quant}}
\end{figure}
\par\end{center}

Figure~(\ref{fig:robust-EU-r2Quant}) compares the simple linear correlation $\overline{R^{2}(\bar{y})}$ to the quantile correlation $\overline{R_{q}^{2}(\bar{y})}$ and to the EU-based \emph{farm-equivalent risk coverage. }The two more sophisticated measures seem to agree remarkably well with the simple linear correlation measure used throughout. The correlation with the quantile pseudo $R^{2}$ is 95\%, while the correlation with the EU-based measure is 84\%. These results are very encouraging, as they suggest that our focus on simple correlation to describe basis risk is justified, since this measure seems to be a strong proxy for more comprehensive measures. Hence,  the bound on basis risk we obtained under the linear correlation criterion may be a good approximation for the bound on these other metrics.

\section{Conclusions\label{sec:Conclusion}}

Basis risk is often recognized as one of the main impediments to the adoption of index insurance. This has led to an extensive literature on the best approach to lowering basis risk. Largely absent in this literature is a recognition that basis risk is comprised of both zonal and design risk, and even the best possible index can only reduce design
risk but not zonal risk. Intuitively, in a region with very heterogeneous fields showing little correlation over time, a good index can do little to reduce basis risk. Without a proper understanding of zonal risk, it is  impossible to distinguish a ``good index in a bad zone`` from a ``bad index in a good zone''. This distinction is, however, crucial since it indicates whether research efforts should be spent on designing better indices or refining insurance zone delineations.  

This paper provides a theoretical decomposition of basis risk and illustrates its application to maize yield data in Kenya. We formalize the basis risk decomposition of \citet{ElabedBellemareEtAl2013}, according to which basis risk can be decomposed within a given insurance zone into zonal and design risk. Ensuring that this decomposition is well-defined requires a notion of the optimal index for a given insurance zone, a question \citet{ElabedBellemareEtAl2013} left open. We fill this gap by deriving the optimal index according to two linear correlation criteria. This provides a useful upper bound on basis risk reduction for a given zone, beyond which basis risk can only be reduced by redefining  the zone itself. 

Our second contribution is to illustrate how this decomposition can be used to understand the sources of basis risk for maize in Kenya. Using satellite-derived maps of maize yields for four years, we investigate the extent of the irreducible zonal risk and of the design risk arising from using a feasible index. We find the irreducible zonal risk to be rather high, about 48\% using the smallest administrative divisions available in Kenya. By definition, this zonal risk can only be reduced by considering refined zones and we investigate the benefits of using satellite data to define smaller zones. The results are rather encouraging, indicating that zonal risk can be reduced to about 40\% using local indices based on 250m neighborhoods around each field. We document also strong regional patterns, with the zonal risk being much higher in the western part of Kenya. Turning to the additional design risk, we find that total basis risk is only slightly increased by using zone-average yields instead of the optimal index, but exacerbated by using weather-based indices such as weather or soil moisture. As a direct consequence, this paper confirms the recommendations by \citet{WB_Clarke_Kenya_2015} to implement an area-yield index scheme in Kenya. Yet, at the same time, our results on the spatial distribution of zonal risk indicate that the success of such a scheme is likely to be limited in some regions.
%Potential solutions to overcome this challenge involve adjusting payout threshold and premium subsidies to the regional context, or the introduction of additional mechanisms such as conditional audits in the case of a mismatch between the index and a farmer's harvest loss \citep{FlatnesCarterEtAl2018}.
Our findings also have important implications for the design of measurement surveys to inform a future area-yield scheme. The strong local heterogeneity we document suggests that accurately measuring average yields will require collecting data on many fields, which could be very costly with traditional approaches. With further advances in remote sensing techniques, the question will soon arise whether using many cheap imperfect measures from satellite data can be more useful than using a few costly ground measures, even if the accuracy of ground measures is higher  \citep[for a related discussion, see][]{BrowningCrossley2009}. 
%A positive answer to this question will open the door to reducing zonal risk by considering smaller zones, which we show here to be potentially very effective. 

This work can be extended in several ways. At the theoretical level,
%a first step toward a theory of basis risk decomposition, 
we focused on linear correlation-based measures of basis risk, which allowed us to derive analytical formulas for the irreducible zonal risk. Arguably, non-linear measures of basis risk based on quantile regression, or even expected utility measures of yields covered by index insurance, are more appropriate to assess the benefits of index insurance. Extending our theoretical results to a risk decomposition based on these criteria will be an beneficial endeavor for future research. Due to the non-linearity of these measures, deriving the optimal index will likely be a challenging task, without a guarantee that an analytical solution for the optimal index exists. Most likely, methods based on numerical approximation will be required, which might become cumbersome for large datasets.
%\footnote{On the other hand, computing the share of the first eigenvalue for very large matrices can be done extremely fast by noting that $X^'X$ and $XX^'$ share the same first eigenvalues and the same trace. As a result}

On the empirical side, our analysis could be extended in several ways, 
notably by using fields instead of pixels and by using longer time series with improved accuracy. 
Doing an analysis at the field level instead of using pixels would reveal important insights into the variability of yields within- and between fields, yet it could be very costly if the data collection were done in the traditional way by manually geo-referencing fields on the ground. Methods based on satellite imagery, either using manual digitization based on crowdsourcing  \citep{EstesEtAl2016} or using  image segmentation based on deep learning \citep{YanRoy2016, WaldnerDiakogiannis2020,WaldnerDiakogiannisEtAl2021} seem very promising. Likewise, the empirical results of this paper were obtained using a relatively short dataset of yields covering just four years. While we believe that this provides essential insights into basis risk in Kenya, reconsideration of our results when longer time series become available will be needed. Given the rapid progress in remote sensing and yield prediction techniques, improved dataset spanning longer time periods should be available in the near future to refine our estimates. This will open the way for an in-depth analysis of our proposal to design local indices with satellite data, which we show here to be potentially very effective at reducing zonal risk. Of particular interest will be to analyze  the trade-off in the optimal zone size between the usefulness of local indices and the risk of moral hazard.

%analyses of longer time periods should be possible in the near future. 

%A second issue pertains to the handling of missing data. In several agricultural systems, crop rotation is practiced regularly, which implies the presence of missing values close to half of the time for a given crop. This complicates the computation of most measures, in particular for eigenvalue problems. More research will be needed to understand how to handle missing values, perhaps with imputation methods. 

%% AJAE
% \printendnotes % AJAE

\bibliographystyle{ecta}%% AJAE: ajae, but error
\bibliography{biblio_mat_synced_from_laptop.bib}

\appendix

\section{Appendix}

%% restart count (cf https://tex.stackexchange.com/questions/85776/change-figure-numbering-for-appendix/85779)
\renewcommand\thefigure{\thesection.\arabic{figure}} 
\renewcommand\thetable{\thesection.\arabic{table}} 
% but hyperref problem, so do: https://tex.stackexchange.com/questions/304814/hyperref-points-wrong-figure-table-with-reset-counter
\renewcommand{\theHtable}{A.\thetable}
\renewcommand{\theHfigure}{A.\thefigure}
\setcounter{figure}{0} 
\setcounter{table}{0} 

\setcounter{thm}{0}
\renewcommand{\thethm}{\Alph{section}.\arabic{thm}}

\subsection{Proofs}

\begin{prop}
\label{proof_opt_vec_R} We can rewrite (\ref{eq:r2_output_gen})
with $\tilde{w}=D^{1/2}w$, and inserting \textup{$D^{-1/2}D^{1/2}$:}
\begin{align}
\mathcal{R}^{2}(w) & =D^{-1/2}\Sigma w\left(w'\Sigma w\right)^{-1}w'\Sigma D^{-1/2}\\
 & =D^{-1/2}\Sigma D^{-1/2}D^{1/2}w\left(w'D^{1/2}D^{-1/2}\Sigma D^{-1/2}D^{1/2}w\right)^{-1}w'D^{1/2}D^{-1/2}\Sigma D^{-1/2}\label{eq:r2_Y_opt_detail}\\
 & =C\tilde{w}\left(\tilde{w}'C\tilde{w}\right)^{-1}\tilde{w}'C\nonumber 
\end{align}

\textup{then Property A.6 (2.3.3) in \citet{Jolliffe2002} shows that
the trace of $\mathcal{R}^{2}(w)$ is maximized using the eigenvector
of C, i.e. $\tilde{w}^{*}=eigenvector(C)$. Then $w^{*}=D^{-1/2}eigenvector(C)=D^{-1/2}v$
is recovered. 
Turning to proposition 1.3: by definition of eigenvector, $Cv=\lambda_{1}v$,
and with normalization $v^{\prime}v=1$ this implies $\lambda_{1}=v'Cv$.
Then $\mathcal{R}^{2}=Cv\left(v'Cv\right)^{-1}v'C=\lambda_{1}v\left(\lambda_{1}\right)^{-1}v'\lambda_{1}=\lambda_{1}vv^{\prime}$.
Now we want $1/N\cdot\Tr(\mathcal{R}^{2})=1/N\cdot\Tr(\lambda_{1}vv^{\prime})=\lambda_{1}/N$.
Because C is a correlation matrix, $\Tr(C)=N=\sum\lambda_j$, hence
$1/N\cdot\Tr(\mathcal{R}^{2})=\lambda_1/\sum\lambda_j$. }
\end{prop}

%\subsection{Supplementary figures}

%\begin{center}
%\begin{figure}
%\caption{Design risk: $\overline{R^{2}}$ from weather \label{fig:design_risk_r2_weather_scatter}}

%\centering{}\includegraphics[width=0.9\columnwidth]{figures/design_risk/r2_wthr_ave_opt}
%\end{figure}
%\par\end{center}

%\clearpage

\end{document}